\documentclass[twocolumn,aps,prl,groupedaddress]{revtex4}
\usepackage{graphicx}
\usepackage{color}
\usepackage{amsmath}
\usepackage{enumitem}

\newcommand{\be}{\begin{equation}}
\newcommand{\ee}{\end{equation}}

\newcommand{\bea}{\begin{eqnarray}}
\newcommand{\eea}{\end{eqnarray}}

\newcommand{\p}{\partial}

\newcommand{\la}{\left\langle}
\newcommand{\ra}{\right\rangle}
\newcommand{\lb}{\left[}
\newcommand{\rb}{\right]}
\newcommand{\lp}{\left(}
\newcommand{\rp}{\right)}
\newcommand{\sgn}{{\rm sgn\,}}

\renewcommand{\Re}{{\rm \, Re\,}}
\renewcommand{\Im}{{\rm \, Im\,}}
\newcommand{\sinc}{{\rm \, sinc\,}}
\renewcommand{\vec}[1]{{\bf #1}}
\newcommand{\bra}[1]{\left\langle #1 \right|}
\newcommand{\ket}[1]{\left|#1\right\rangle}
\newcommand{\braket}[2]{\left\langle#1 |  #2\right\rangle}

\newcommand{\ketbra}[2]{| #1 \rangle\langle #2 |}
\newcommand{\rd}[1]{\mathop{\mathrm{d}#1}}

\begin{document}
\title{Higher-Than-Ballistic Conduction of Viscous Electron Flows}

\author{Haoyu Guo$^1$, Ekin Ilseven$^1$, Gregory Falkovich$^{2}$ and Leonid Levitov$^1$}

\affiliation{$^1$Massachusetts Institute of Technology, Cambridge, Massachusetts 02139, USA \\ $^2$Weizmann Institute of Science, Rehovot 76100 Israel}


\begin{abstract}
 Strongly interacting electrons can move in a neatly coordinated way, reminiscent of the movement of viscous fluids. Here we show that in viscous flows interactions facilitate transport,
allowing conductance to exceed the fundamental Landauer's ballistic limit $G_{\rm ball}$. The effect is particularly striking for the flow through a viscous point contact, a constriction exhibiting the quantum-mechanical ballistic transport at $T=0$ but governed by electron hydrodynamics at elevated temperatures. We develop a theory of
the ballistic-to-viscous crossover using an approach based on quasi-hydrodynamic variables. Conductance is found to obey an additive relation $G=G_{\rm ball}+G_{\rm vis}$, where the viscous contribution $G_{\rm vis}$
dominates over $G_{\rm ball}$ in the hydrodynamic limit. 
We argue that superballistic, low-dissipation transport is a generic feature 
of viscous electronics.
\end{abstract}

\maketitle

Free electron flow through constrictions in metals is often regarded as an ultimate high-fidelity charge transfer \cite{sharvin65,van_wees_88,wharam_88,maslov95,van_houten_96}. Can conductance ever exceed the ballistic-limit value?
Here we show that this is possible for strongly interacting systems in which electron movement resembles that of viscous fluids. Electron fluids are predicted to occur in quantum-critical systems and in high-mobility conductors, so long as momentum-conserving electron-electron scattering dominates over 
other scattering processes\cite{gurzhi63,jaggi91,LifshitzPitaevsky_Kinetics,damle97}. 
Viscous electron flows feature a host of novel transport behaviors\cite{refB1,refB3,andreev2011,sheehy2007,fritz2008,muller2009,mendoza2011,forcella2014,tomadin2014,narozhny2015,principi2015,cortijo2015,lucas2016}.  Signatures of such flows have been
observed in
ultra-clean GaAs, graphene and ultrapure PdCoO${}_2$ \cite{dejong_molenkamp,bandurin2015,crossno2016,moll2016}.

\begin{figure}
\includegraphics[width=0.42\textwidth]{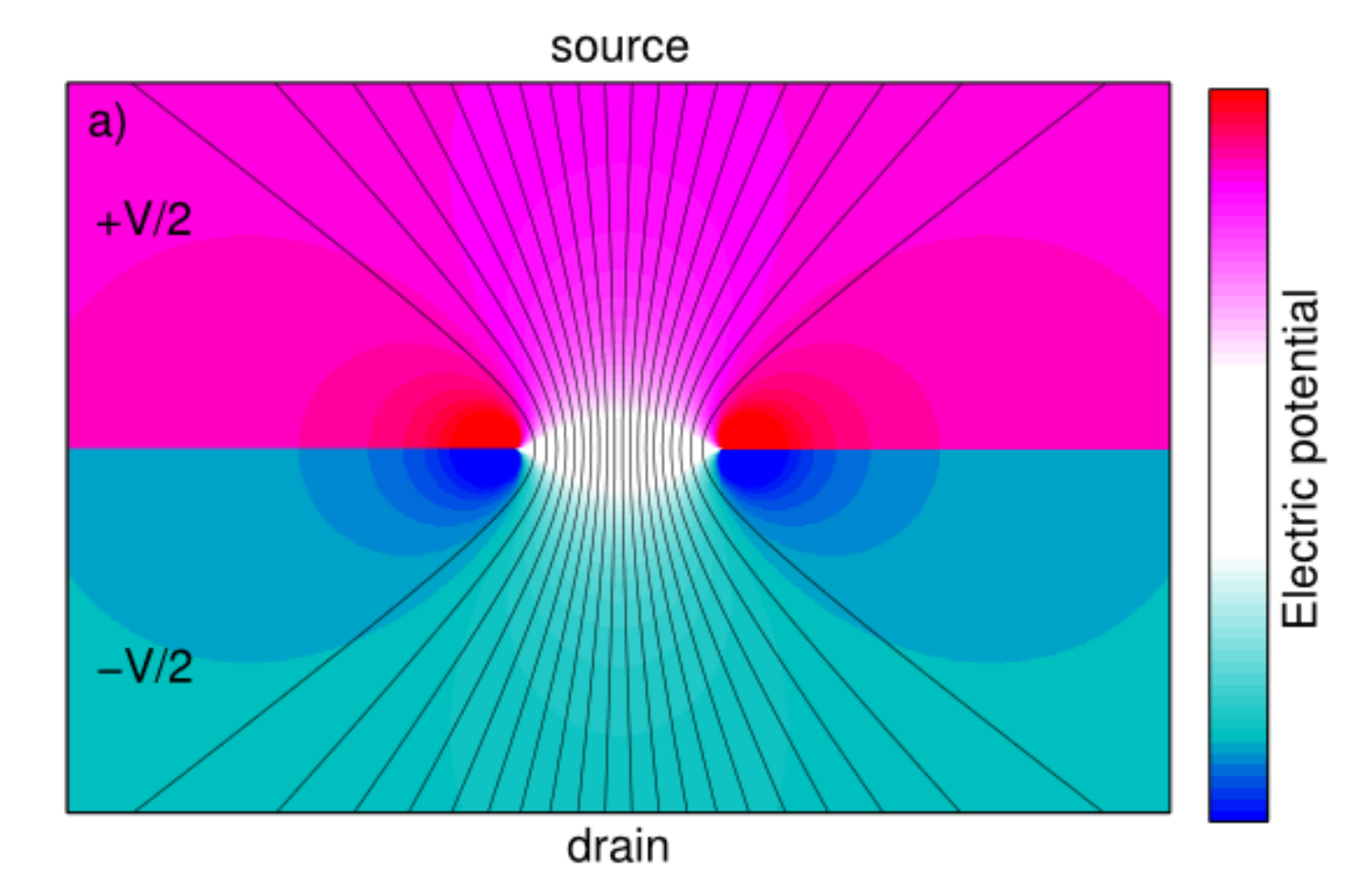} 
\includegraphics[width=0.41\textwidth]{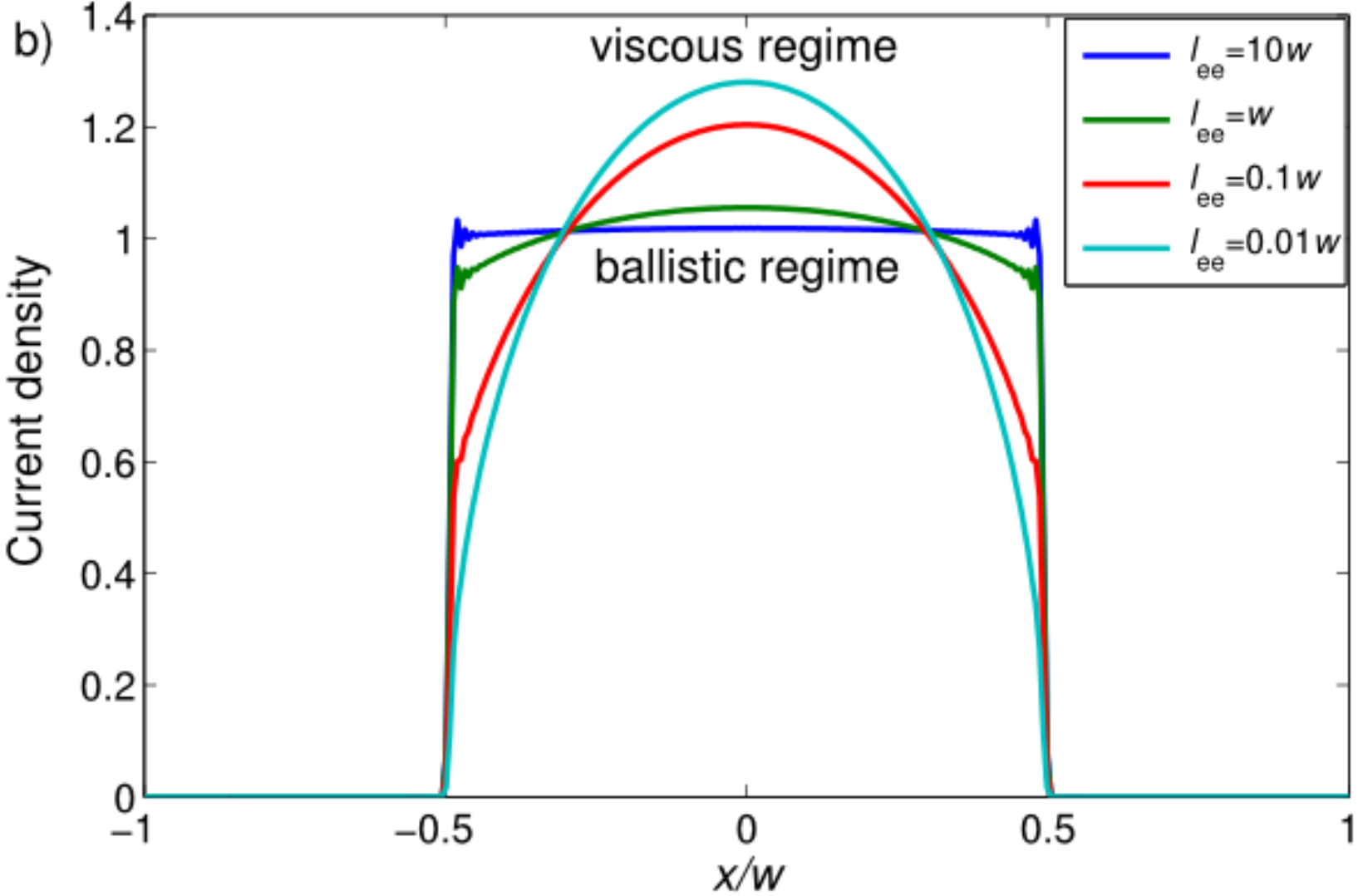}
\caption{ a) Current streamlines (black) and potential colormap for viscous flow through a constriction. 
Velocity magnitude is proportional to the density of streamlines. 
Current forms a narrow stream, 
avoiding the boundaries where dissipation occurs and allowing the resistance, Eq.\eqref{eq:R_vis}, to drop below the ballistic-limit value.
b) Current distribution in the constriction
for different carrier collision 
mean-free-path values. 
The distribution evolves from a constant in the ballistic regime 
to a semicircle in the viscous regime, 
Eq.\eqref{eq:semicircle}, illustrating the interaction-induced streaming effect. 
Parameters used: $L=3w$, $b=10^5v$. A Fourier-space filter was used to smooth out the Gibbs phenomenon. 
}
\label{fig1}
\vspace{-5mm}
\end{figure}

We will see that electrons in a viscous flow can achieve through cooperation what they cannot accomplish individually.
As a result, resistance and dissipation of a viscous flow
can be markedly smaller than that for the free-fermion transport. As a simplest realization, we discuss  {\it viscous point contact} (VPC) where
correlations act as a `lubricant' facilitating the flow. The reduction in resistance arises due to the streaming effect illustrated in Fig.\ref{fig1}, wherein electron currents bundle up to form 
streams that bypass the boundaries, where momentum loss occurs.
This surprising behavior is in a clear departure from the common view that regards electron interactions as an impediment for transport.

A simplest VPC is a two-dimensional constriction pictured in Fig.\ref{fig1}a.
The interaction effects dominate 
in constrictions of width $w$ exceeding 
the carrier  collision 
mean free path ${l_{\rm ee}}$ (and much greater than the Fermi wavelength $\lambda_F$).
The VPC conductance, evaluated 
in the absence of impurity scattering, scales as a square of the 
width $w$ and inversely with the electron viscosity $\eta$:
\be\label{eq:R_vis}
G_{\rm vis}(w)= 
\frac{\pi n^2e^2 w^2}{32\eta}
%
,\quad w\gg{l_{\rm ee}}
,
\ee
where $n$ and $e$ are the carrier density and charge.
In the opposite limit, ${l_{\rm ee}}\gg w$, the ballistic free-fermion model \cite{sharvin65,van_houten_96} predicts the conductance
$G_{\rm ball}=\frac{2e^2}{h}N $,
where $N\approx 2w/\lambda_F$ is the number of Landauer's open transmission channels.
The conductance 
$G_{\rm vis}$ grows with width 
faster than $G_{\rm ball}$.
Therefore, for large enough $w$, viscous transport 
yields $G$ values above the ballistic bound. 

Conveniently, both regimes are accessible in a single constriction, since transport is expected to be viscous at elevated temperatures and ballistic at $T=0$. The crossover temperature can be estimated in terms of the electron-electron scattering mean free path as
\be\label{eq:l/w=0.3}
{l_{\rm ee}}(T)/w=\pi^2/16\approx 0.62
.
\ee
This relation is found by setting $R_{\rm vis}=R_{\rm ball}$ and writing $\eta=\nu nm=\frac14v_F{l_{\rm ee}}nm$, with $m$ the carrier mass
and the kinetic viscosity $\nu$ estimated in Eq.\eqref{eq:viscous_mode}. 
The condition \eqref{eq:l/w=0.3} can be readily met in micron-size graphene junctions.

Several effects of electron interactions on transport in constrictions 
were discussed recently. Refs.\cite{refB1,refB3} study junctions with  spatially varying electron density and, using the time-dependent current-density functional theory,  predict a suppression of conductance. A hydrodynamic picture of this effect was established in Ref.\cite{andreev2011}.  In contrast, here we study junctions in which, in the absence of applied current, the carrier density is approximately position-independent. This situation was analyzed in Ref.\cite{nagaev08} perturbatively in the ee scattering rate, finding a conductance enhacement that resembles our results.

The relation \eqref{eq:R_vis} points to a simple way to measure viscosity by the conventional transport techniques. Precision measurements of viscosity in fluids date as far back as 19 century\cite{marvin71}.
They relied, in particular, on measuring resistance of a viscous fluid discharged through a narrow channel or an orifice, 
a direct analog of our constriction geometry. Further, 
viscosity-induced electric conduction 
has a well known counterpart in the kinetics of classical gases, where momentum exchange between atoms results in a slower momentum loss and a lower resistance of gas flow. It is responsible, in particular, for a dramatic drop in the hydrodynamic resistance upon a transition from Knudsen 
to Poiseuille regime. 
For a viscous flow through scatterers spaced by a distance $L$ the typical time of momentum transfer is $\tau\sim L^2/\nu\sim L^2/v_T\ell$, whereas for an ideal gas this time is $\tau'=L/v_T$, where $v_T$ is thermal velocity and $\ell$ is the mean free path. For $\ell\ll L$ the viscous time $\tau$ is much longer than the ballistic time $\tau'$.

The peculiar correlations originating from fast particle collisions in proximity to  scatterers
can be elucidated by a spacial argument: particle collisions near a scatterer reduce the average velocity component normal to the scatterer surface, $v_\perp$, which slows down the momentum loss rate per particle, $m\vec v\,v_\perp /L$. Momentum exchange makes particles flow collectively, on average staying away from scatterers and thus lowering the resistance.

The viscosity-induced drop in resistance 
can be used
as a vehicle to overcome the quantum-ballistic limit for electron conduction.
Indeed, we can compare the values $R_{\rm vis}$ and $R_{\rm ball}$ by putting 
them in a Drude-like form $R=m/ne^2\tau$, with $m$ the carrier mass and $\tau$ a suitable momentum relaxation time. Eq.\eqref{eq:R_vis} can be modeled in this way using the time of momentum diffusion across the constriction $\tau=w^2/\nu$, whereas $R_{\rm ball}$
can be put in a similar form with $\tau'=w/v_F$ the flight time across the constriction. Estimating $\nu=\frac14 v_F{l_{\rm ee}}$, we see that Eq.\eqref{eq:R_vis} predicts resistance below the 
ballistic-limit values so long as $\tau\gtrsim\tau'$, i.e. in the hydrodynamic regime $w\gtrsim{l_{\rm ee}}$.

Understanding the behavior at the ballistic-to-viscous crossover 
is a nontrivial task. Here, to tackle the 
crossover, we use kinetic equation with a simplified 
ee collision operator 
chosen in such a way that the relaxation rates for all nonconserved harmonics of 
momentum distribution
are the same. 
This model provides a closed-form solution for transport through VPC 
for any ratio of 
the lengthscales $w$ and ${l_{\rm ee}}$, predicting 
a remarkably simple additive relation
\be\label{eq:G=G+G}
G_{\rm VPC}=G_{\rm ball}+G_{\rm vis}
.
\ee
This dependence, derived from a microscopic model,  
interpolates between 
the ballistic and viscous limits, $w\ll {l_{\rm ee}}$ and $w\gg {l_{\rm ee}}$,
in which the terms $G_{\rm ball}$ and $G_{\rm vis}$ 
dominate, respectively.

We start with a simple derivation of the VPC 
resistance in Eq.\eqref{eq:R_vis} using the model of a low-Reynolds electron flow that obeys the Stokes equation \cite{LF}.
\be
(\eta\nabla^2-(ne)^2\rho)\vec v(\vec r)=ne\nabla\phi(\vec r)
.
\label{eq:stokes}
\ee
Here $\phi(\vec r)$ is the electric potential, $\eta$ is the viscosity and the second term
describes ohmic resistivity due to impurity or phonon scattering.
Our analysis relies on a symmetry argument and invokes an auxiliary electrostatic problem. We model the constriction in Fig.\ref{fig1}a as a slit $-{\frac{w}2}<x<{\frac{w}2}$, $y=0$. 
The $y\to -y$ symmetry ensures that the current component $j_y$ is an even function of $y$ whereas both the component $j_x$ and the potential $\phi$ are odd in $y$. As a result the quantities
$j_x$ and $\phi$ vanish within the slit at $y=0$. 
This observation allows us to write the potential in the plane as a superposition of contributions due to different current elements in the slit
\be\label{eq:phi_general}
\phi(x,y)=\int_{-{\frac{w}2}}^{{\frac{w}2}} dx' {\cal R}(x-x',y) j(x')
,
\ee
where the influence function ${\cal R}(x,y)=\frac{\beta (y^2-x^2)}{(x^2+y^2)^2} $ 
describes potential in a halfplane due to a point-like current source at the edge, obtained from 
Eq.\eqref{eq:stokes} with no-slip boundary conditions and $\rho=0$ \cite{FL}.
Here $\beta={ 2\eta\over \pi(en)^2}$ and without loss of generality we focus on the $y>0$ halfplane.

Crucially, rather than providing a solution to our problem, the potential-current relation \eqref{eq:phi_general} merely helps to pose it. Indeed, a generic current distribution would yield a potential which is not constant inside the slit. We must therefore determine the functions $j(x)$ and $\phi(x,y)$ self-consistently, in a way that ensures that the resulting $\phi(x,y)$ vanishes on the line $y=0$ inside the slit. 
Namely, Eq.\eqref{eq:phi_general} must be treated as an integral equation for an unknown function $j(x)$.
Denoting potential values at the halfplane $y\ge 0$ edge as $\phi_{+0}(x)=\phi(x,y)_{y=+0}$, we can write the relation \eqref{eq:phi_general} as
\be\label{eq:int_eqn}
\phi_{+0}(x)=-\frac{\beta}2\int_{-\infty}^\infty dx' \lp \frac{j(x')}{(x-x'+i0)^2}+{\rm c.c.}\rp
,
\ee
where $j(x)$ is the current $y$ component, which is finite inside and zero outside the interval $[-{\frac{w}2},{\frac{w}2}]$.

A solution of this integral equation such that $\phi_{+0}(x)$ vanishes for all $-{\frac{w}2}<x<{\frac{w}2}$  can be obtained from a 3D electrostatic problem for an ideal-metal strip of width $w$ placed in a uniform external electric field $\vec E_0=\lambda \hat{\vec x}$. The strip is taken to be infinite, zero-thickness, and positioned in the $Y=0$ plane such that
\be
{\textstyle 
-{\frac{w}2}<X<{\frac{w}2},\quad Y=0,\quad -\infty<Z<\infty
}
\ee
 (for clarity we denote 3D coordinates by capital letters). Potential $\Phi_{3D}(X,Y)$ is a harmonic function,  constant on the strip and behaving asymptotically as  $-E_0 X$. It is easily checked that the 3D electrostatic problem translates to the 2D viscous problem as
\be\label{eq:3D2Dmapping}
\begin{array}{lccc}
3D, \, Y=0: & X & \sigma(X) & E_x(X)
\\
 & \downarrow & \downarrow & \downarrow
 \\
2D, \, y=+0: & x & 
-\frac{\beta}{2}\p j/\p x & \phi_{+0}(x)
\end{array}
.
\ee
This mapping transforms Coulomb's charge-field relation between the electric field at $Y=0$ and the surface charge density,
$E_x(X)=2\int_{-\infty}^\infty \frac{\sigma(X')dX'}{X-X'}$, into the 2D viscous relation in Eq.\eqref{eq:int_eqn}. Potential $\Phi_{3D}$, obtained through a textbook application of conformal mapping, then equals
\be\label{eq:phi_strip}
{\textstyle
\Phi_{3D}(X,Y)=-\Re \lambda \sqrt{ \zeta^2-{\frac{w^2}4}} 
,\quad \zeta=X+iY
.}
\ee
%
Eq.\eqref{eq:phi_strip} describes the net contribution of the external field $\vec E_0$ and the charges $\sigma(X)$ induced on the strip. 
The field component $E_x(X)=-\p_X \Phi_{3D}$ vanishes on the strip $-{\frac{w}2} < x < {\frac{w}2}$ and equals $\lambda$ far outside.
We can therefore identify $\lambda$ with $V/2$ in the viscous problem (see Fig.\ref{fig1}a).

Charge density on the strip, found from (\ref{eq:phi_strip}) with the help of Gauss' law, $\sigma(X) 
=\frac{\lambda X}{2\pi\sqrt{w^2/4-X^2}} $, under the mapping \eqref{eq:3D2Dmapping} gives a semicircle current distribution:
\be\label{eq:semicircle}
{\textstyle
j(|x|<{\frac{w}2} 
)=\frac{\lambda}{\pi\beta}  \sqrt{{\frac{w^2}4}-x^2}
,\quad
j(|x|>{\frac{w}2} )=0
.}
\ee
Potential map in Fig.\ref{fig1}a is then obtained by plugging this result in Eq.\eqref{eq:phi_general}. The flow streamlines are obtained from a  similar relation for the stream function, see \cite{FL}. Evaluating the current $I=\int_{-{\frac{w}2}}^{\frac{w}2} j(x)dx=\lambda w^2/8\beta$ and setting $\lambda=V/2$ 
yields $R=V/I=16\beta/w^2$
which is Eq.\eqref{eq:R_vis}.
The inverse-square scaling $R\propto w^{-2}$ is distinct from the $w^{-1}$ scaling  found in the ballistic free-fermion regime. The scaling, as well as the lower-than-ballistic $R$ values, can serve as a hallmark of a viscous flow.

Potential, inferred from the 2D/3D correspondence, is
\be\label{eq:phi_viscous_y=0}
\phi(x)_{y=\pm 0}=\left\{\begin{array}{cc} \frac{ V|x|}{2\sqrt{x^2-w^2/4}}\sgn y & |x|>{\frac{w}2} \\ 0 & |x|<{\frac{w}2}
\end{array}\right.
\ee 
where $\sgn y$ corresponds to the 
the upper and lower sides, $y=\pm 0$. 
Potential grows towards the slit, diverging at the end points $x=\pm{\frac{w}2}$. This interesting  behavior, representing an up-converting DC-current transformer, arises due to the electric field pointing against the current near the viscous fluid edge\cite{LF}.

 \begin{figure}
  \centering
  \includegraphics[width=0.4\textwidth]{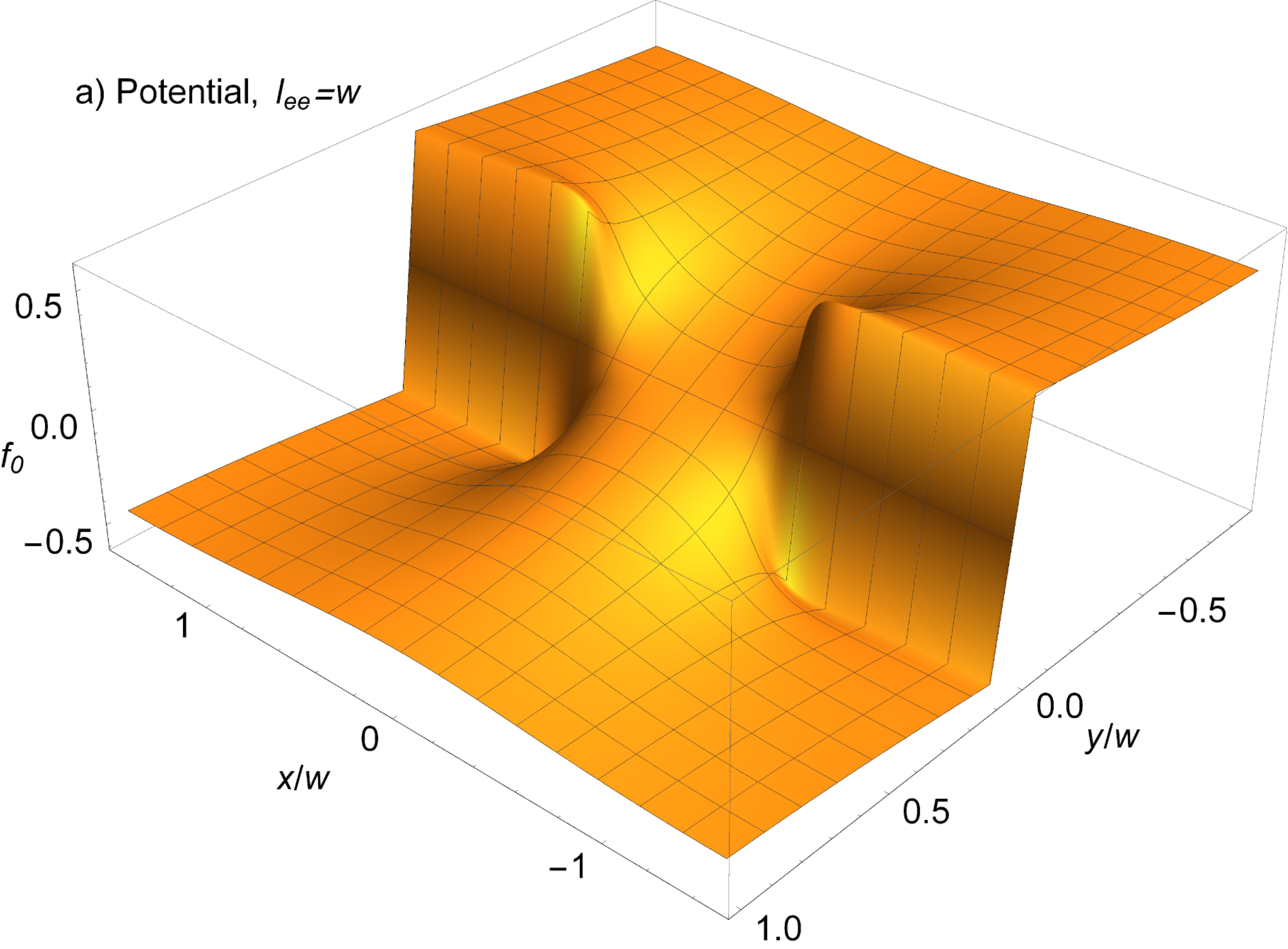} 
    \includegraphics[width=0.4\textwidth]{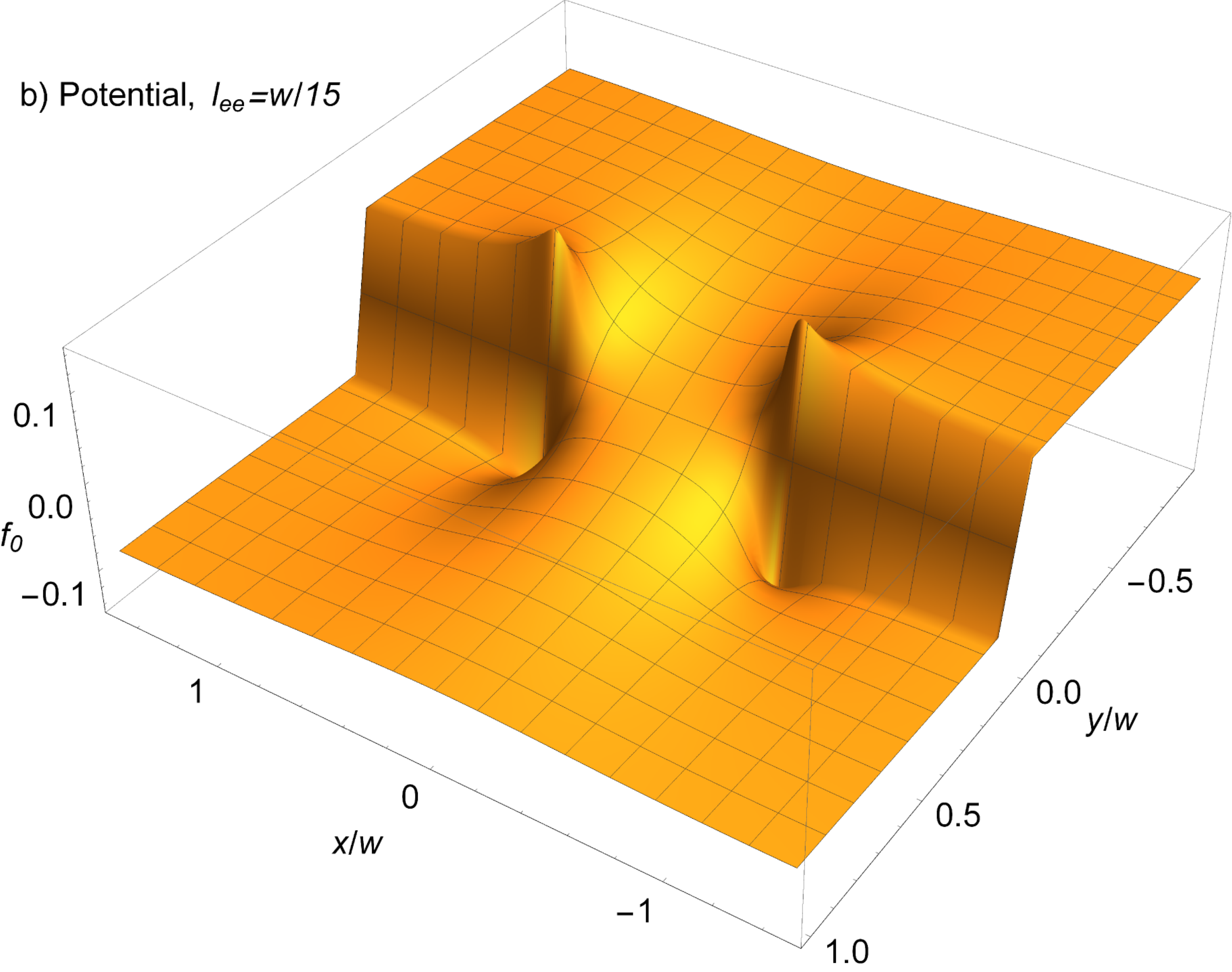} 
  \caption{Potential distribution induced by current through a constriction (a) at the crossover, ${l_{\rm ee}}\sim w$, and (b) in the viscous regime, ${l_{\rm ee}}\gg w$. The spikes at the constriction edges in b) is a signature of a hydrodynamic behavior, see Eq.\eqref{eq:phi_viscous_y=0} and accompanying text.  Plotted is particle density deviation from equilibrium, $f_0(\vec x)$, which is proportional to potential (see text). 
  Parameters used:
 (a)  $\gamma=v/w$,  (b) $\gamma=15v/w$; other parameter values are the same as in Fig.\ref{fig1}b.
}\label{fig2}
\end{figure}

Our next goal is to develop a theory of the  ballistic-to-viscous crossover 
for a constriction.
Since we are interested in the linear response, we use the kinetic equation
linearized in deviations of particle distribution from the equilibrium Fermi step (assuming $k_{\rm B}T\ll E_F$),
%
\be\label{eq:Boltzmann}
\lp\p_t+\vec v\nabla_{\vec x}\rp f(\theta,\vec x,t)=I_{\rm ee}(f)+I_{\rm bd}(f)
,
\ee
where $\theta$ is the angle parameterizing particle momentum at the 2D Fermi surface. 
Here $I_{\rm ee}$ and $I_{\rm bd}$
describe momentum-conserving carrier 
collisions 
and momentum-nonconserving scattering at the boundary, respectively.


In the presence of
momentum-conserving collisions transport is 
succinctly described 
by {\it quasi-hydrodynamic variables} defined as deviations in the average
particle density and momentum from local equilibrium\cite{levinson77}. These quantities can be expressed
 as angular harmonics of the distribution $f(\theta,\vec x,t)$:
\be\label{eq:zero_modes}
f_0=\la f(\theta)\ra_\theta
,\quad
f_{\pm 1}=\la e^{\mp i\theta}f(\theta)\ra_\theta
\ee
where we introduced notation $\la...\ra_\theta= \oint ...\frac{d\theta}{2\pi}$. The quantities
$f_0$, $f_{\pm 1}$, conserved in the ee collisions,
represent the zero modes of $I_{\rm ee}$. For suitably chosen  $I_{\rm ee}$ the task of solving the kinetic equation 
in a relatively complicated constriction geometry is reduced to analyzing a selfconsistency equation for the 
variables $f_0$, $f_{\pm 1}$. 
We will derive a linear integral equation for these quantities, 
and solve it to obtain the current density, potential and conductance.

To facilitate the analysis, we model $I_{\rm ee}$ 
by choosing a single relaxation rate for all non-conserved harmonics:
\be\label{eq:Iee_Idis}
I_{\rm ee}(f)=-\gamma(f-Pf)
,\quad
P= 
\sum_{m=0,\pm 1} \left. |m\ra \la m|\right.
,
\ee
%
where $\gamma$ represents the ee collision rate, with ${l_{\rm ee}}=v/\gamma$,
and $P$ is a projector in the space of angular harmonics of $f(\theta)$ that selects the harmonics conserved in ee collisions. Here we introduced Dirac notation for 
$f(\theta)$ with the inner product $\la f_1|f_2\ra =\oint\frac{d\theta}{2\pi} \bar f_1(\theta)f_2(\theta)$. Namely,
\be\nonumber
\la\theta|m\ra=e^{im\theta}
,\quad
Pf(\theta)=\sum_{m=0,\pm1}
\oint\frac{d\theta'}{2\pi}  e^{im(\theta-\theta')}f(\theta') 
.
\ee
As in quantum theory, the Dirac notation proves to be a useful bookkeeping tool to account on equal footing for the distribution function position and wavenumber dependence, as well as the angle dependence. 

To simplify our analysis we replace the constriction geometry by that of a full plane, with a part of the line $y=0$ made impenetrable through a suitable choice of $I_{\rm bd}(f)$. 
Scattering by disorder at the actual boundary conserves $f_0$ but not $f_{\pm 1}$.
We can therefore model momentum loss due to collisions at the boundary 
using
\be\label{eq:alpha(x)}
I_{\rm bd}(f)=-\alpha(\vec x)P'f
,\quad
\alpha(\vec x)=\left\{\begin{array}{cc} 0, & |x|<{\frac{w}2} \\
b\delta(y), & |x|\ge {\frac{w}2}
\end{array}\right.
\ee
where $P'$ is a projector defined in a manner similar to $P$,  projecting $f$ on the harmonics $m=\pm1$.
The term $\alpha(x)$ describes momentum relaxation on the line $y=0$, equal zero within the slit and $b$ outside.
The parameter 
$b>0$ with the dimension of velocity, introduced for mathematical convenience, describes partially transparent boundary. 
An impenetrable no-slip boundary, which corresponds to the situation of interest, can be modeled by taking the limit $b\to\infty$.

We will analyze the flow 
induced by a current applied along the $y$ direction, described by a distribution 
\be
f(\theta,\vec x)=f^{(0)}(\theta)
+\delta f(\theta,\vec x)
,\quad f^{(0)}(\theta)\sim \sin\theta
.
\ee 
Here $f^{(0)}$ and $\delta f$, which we will also write as $\ket{ f^{(0)}}$ and $\ket{\delta f}$, 
represent a uniform  current-carrying state and its distortion due to scattering at the $y=0$ boundary.  
Once found, the spatial distribution $f(\theta,\vec x)$ 
will allow us to determine the resulting potential and resistance. 
The kinetic equation, Eq.\eqref{eq:Boltzmann}, reads
\be\label{eq:Boltzmann_J0}
(\p_t+K+\alpha(\vec x)P')\ket{f}=0
,\quad
K=\vec v\nabla+\gamma \hat 1-\gamma P
\ee
(from now on we suppress the coordinate and angle dependence of $f$ 
and use the Dirac notation). Plugging $f=f^{(0)}+\delta f$,
we rewrite Eq.\eqref{eq:Boltzmann_J0} as
$
\lp 
K+\hat\alpha \rp \ket{\delta f}
=-\hat\alpha\ket{f^{(0)}}
$,
where,  
for conciseness, we absorbed the projector $P'$ into $\hat\alpha$ and set $\p_t f=0$ for a steady state. We write a formal operator solution as 
\be\label{eq:df}
\ket{ \delta f}=-(1+G\hat\alpha)^{-1}G\hat\alpha \ket{f^{(0)}}
\ee
where $G=K^{-1}$ 
is the Greens function.
Performing analysis in momentum representation, we treat the scattering term in Eq.\eqref{eq:alpha(x)} as an operator
\be
\la \vec k|\hat\alpha|\vec k'\ra=P' \alpha_{k_1-k'_1}
, \quad
\alpha_k=2\pi b\delta(k)-bw\sinc\frac{kw}{2}
\ee
where $\sinc x=\frac{\sin x}{x}$. The two terms in $\alpha_k$
describe scattering at the $y=0$ line less the slit contribution. 

Next we derive a closed-form integral equation for quasi-hydrodynamic variables. This is done by projecting the quantities in Eq.\eqref{eq:df}
on the 
$m=0,\pm1$ harmonics, Eq.\eqref{eq:zero_modes}.
Acting on 
Eq.\eqref{eq:df} with $P$ gives 
%
$
\ket{P \delta f}= -(1+\tilde G\hat\alpha)^{-1}\tilde G\hat\alpha \ket{f^{(0)}}
$
where
$\tilde G=PGP$ is a $3\times3$ matrix in the $m=0,\pm1$ space (here
we used the identity $\hat\alpha=P\hat\alpha P$ which follows from $PP'=P'P=P'$). 
The integral equation is obtained by acting on both sides with the operator $1+\tilde G \hat\alpha$, giving
\be\label{eq:(1+Ga)f=f0}
(1+\tilde G\hat\alpha)\ket{\tilde f}=\ket{f^{(0)}}
.
\ee
Here we defined $\tilde f=f^{(0)}+P \delta f$, the full distribution function projected on the $m=0,\pm1$ harmonics. 

The quantity $\tilde f$  represents an unknown function which can be found, in principle, by inverting the integral operator $1+\tilde G\alpha$ in Eq.\eqref{eq:(1+Ga)f=f0}. However, rather than attempting to invert $1+\tilde G\alpha$ directly in 2D, it is more convenient to proceed in two steps: first analyze Eq.\eqref{eq:(1+Ga)f=f0} in 1D, on the line $y=0$,  and then extend the solution into 2D. 

We start with finding $\tilde G$. 
As a first step, we evaluate the $3\times 3$ matrix $S=\gamma PG_0P$ where $G_0=1/(i\vec k\vec v+\gamma)$. The quantity $G_0$ is an auxiliary Greens function describing transport in which all harmonics, including $m=0,\pm1$, relax at a rate $\gamma$. Direct calculation gives matrix elements (here $m,m'=0,\pm 1$, $\Delta m=m'-m$):
\be\label{eq:g3x3}
S_{mm'}=\la \frac{\gamma e^{i(m'-m)\theta}}{\gamma+i \vec k \vec v}\ra_\theta
\!\!
=\tanh\beta \frac{ e^{i\theta_k\Delta m} }{\lp ie^{\beta}\rp^{|\Delta m|}}
,
\ee
where we denote $\sinh\beta=\frac{\gamma}{kv}$ and $\theta_k=\arg(k_1+ik_2)$.

The matrix $\tilde G$ can now be expressed through the matrix $S$
by expanding the actual Greens function as $G=1/(G_0^{-1}-\gamma P)=G_0+G_0\gamma PG_0+...$, which gives
\be\label{eq:G_Tmatrix}
G=G_0+G_0TG_0
,\quad
T=\frac{\gamma P}{1-\gamma PG_0P}
.
\ee
Here we re-summed the series, expressing the result in terms of a $3\times 3$ matrix $T$ in a manner analogous to the derivation of the Lippmann-Schwinger $T$-matrix for quantum scattering with a finite number of `active' channels.
We note that $\gamma PG_0P$ is nothing but the matrix $S$ in Eq.\eqref{eq:g3x3}. 
Plugging 
Eq.\eqref{eq:G_Tmatrix} into $\tilde G=PGP$
and performing a tedious but straightforward 
matrix inversion we obtain 
\be 
\tilde G=\frac{\gamma^{-1}S}{1-S}
=\frac{\sinh\beta}{\gamma} \!\lp\begin{array}{ccc} e^{\beta} & -i \bar z_k &-e^{\beta} \bar z_k^2\\ -i z_k & e^{-\beta} & -i \bar z_k \\ -e^{\beta}  z_k^2 & -i  z_k & e^{\beta} \end{array}\rp
,
\ee
where $z_k=e^{i\theta_k}$ 
and the basis vectors are ordered as $\ket{+1}$, $\ket{0}$, $\ket{-1}$.

In what follows it will be convenient to transform $\ket{\pm 1}$ to the even/odd basis 
$\ket{c}=\frac{\ket{+1}+\ket{-1}}{\sqrt{2}}$, $\ket{s}=\frac{\ket{+1}-\ket{-1}}{\sqrt{2}i}$. In this basis $\tilde G$ reads
\be 
\label{eq:Gtilde}
\lp\begin{array}{ccc} \tilde G_{00} & \tilde G_{0c} & \tilde G_{0s} \\ \tilde G_{c0} & \tilde G_{cc} & \tilde G_{cs} \\ \tilde G_{s0} & \tilde G_{sc} & \tilde G_{ss}\end{array}\rp
=\left(
\begin{array}{ccc}
 \frac{R_-}{\gamma  \kappa^2} & \frac{-i \sqrt{2} \kappa_1}{\gamma \kappa^2} &
   \frac{-i\sqrt{2} \kappa_2}{\gamma \kappa^2} \\
 \frac{- i \sqrt{2} \kappa_1}{\gamma \kappa^2} & \frac{2  \kappa_2^2
   R_+}{\gamma  \kappa^4} & \frac{-2  \kappa_1  \kappa_2 R_+}{\gamma   \kappa^4} \\
 \frac{-i \sqrt{2} \kappa_2}{\gamma \kappa^2 } & \frac{-2  \kappa_1  \kappa_2
  R_+}{\gamma  \kappa^4} & \frac{2  \kappa_1^2 R_+}{\gamma  \kappa^4} \\
\end{array}
\right)
,
\ee
where the basis vectors are ordered as  $\ket{0}$, $\ket{c}$, $\ket{s}$ and
we defined $R_{\pm}(\kappa)=\sqrt{ \kappa^2+1}\pm 1$ and $ \kappa_{1,2}=\frac{v}{\gamma}k_{1,2}$, $\kappa=\sqrt{\kappa_1^2+\kappa_2^2}$. 
The quantities $G$ and $\tilde G$ represent, through their dependence on $\vec k$,  translationally invariant integral operators in position representation and diagonal operators in momentum representation.

Next, we evaluate the matrix that represents the operator $\tilde G$ restricted to the line $y=0$,
\be\label{eq:F=int D}
D(k_1)=\int_{-\infty}^{\infty} \frac{dk_2}{2\pi} \tilde G(k_1,k_2)
.
\ee 
The matrix elements $\tilde G_{0c}$ and  $\tilde G_{0s}$ are odd in $k_2$ and therefore give zero upon integration in Eq.\eqref{eq:F=int D}. 
This gives a block-diagonal matrix
\be
D(k_1)=\lp\begin{array}{ccc} D_{00}(k_1) & D_{0c}(k_1) & 0 \\ D_{c0}(k_1) & D_{cc}(k_1) & 0 \\ 0 & 0 & D_{ss}(k_1)\end{array}\rp
.
\ee
The quantity $D_{ss}(k_1)$ will play a central role in our analysis. 
Indeed, since the flow of interest is symmetric under $y\to -y$ and $x\to -x$, 
the $\tilde f_0$ and $\tilde f_c$ components vanish on the $y=0$ line. 
As a result, the distribution function at $y=0$ is of a pure $\ket{s}$ form  i.e. $\tilde f(\theta,x)=g(x)\sqrt{2}\sin\theta$.

Evaluating the integral over $k_2$ in Eq.\eqref{eq:F=int D} we obtain
\be\label{eq:Dss(k)}
D_{ss}(k)=\frac{\frac{\pi}2  \sgn \kappa + \kappa+(\kappa^2+1) \cot^{-1}\kappa  }{\pi  \kappa v}
\ee
where $\kappa=kv/\gamma$.
This expression defines an even function of $k$ with the asymptotics 
\be\label{eq:Fss_asymptotic}
D_{ss}(|k|v\ll \gamma)=\frac{\gamma}{|k|v^2}
,\quad
D_{ss}(|k|v\gg \gamma)=\frac2{\pi v}
.
\ee
Since the matrix element $D_{ss}$ is an eigenvalue of $D$ for 
the eigenvector $\ket{s}$,  the $\theta$ dependence can be factored out of Eq.\eqref{eq:(1+Ga)f=f0},  giving 
$(1+D\alpha)\ket{g}=\ket{g^{(0)}}$. 
Finally, multiplying 
by $D^{-1}$, we obtain the `central equation' 
\be\label{eq:line_y=0}
D_{ss}^{-1}(k) g_k+\int\frac{dk'}{2\pi}\alpha_{k-k'}g_{k'}=2\pi \mu\delta(k)
,
\ee
where $\mu$ is an unspecified number, akin to a Lagrange multiplier, which fixes the total current value. 
Here, we wrote the relation $(D^{-1}+\alpha)\ket{g}=\mu\ket{k=0}$ as an integral equation, replacing $k_1$ with $k$ for clarity.

The origin of the $\mu$-term in \eqref{eq:line_y=0}, and its relation with the properties of the operator $D$, is simplest to understand using a discretized momentum representation. 
Letting $k_1=\frac{2\pi}{L}n$ and replacing 
\be\label{eq:discrete_k}
\int  dk_1 ...\to\frac{2\pi}{L}\sum_n ...\, 
,\quad
2\pi \delta(k)\to L\delta_{k,0}
,
\ee 
i.e. putting the problem on a cylinder of circumference $L$, 
we see that the values  $\tilde G_{ss}(k_1,k_2)$ vanish for $k_1=0$ and any $k_2$. This means that $D_{ss}(k_1)$ also vanishes for $k_1=0$ and thus the operator $D$ does not have an inverse. In this case caution must be exercised when multiplying by $D^{-1}$. Namely, the quantities $D^{-1}\ket{f}$ are defined modulo a null vector of $D$, which is the $k_1=0$ mode with an unspecified coefficient, represented by the $\mu$-term. We note parenthetically that discretization has no impact on the values $D_{ss}(k_1\ne 0)$ given in Eqs.\eqref{eq:Dss(k)},\eqref{eq:Fss_asymptotic}.

     \begin{figure} 
  \includegraphics[width=0.4\textwidth]{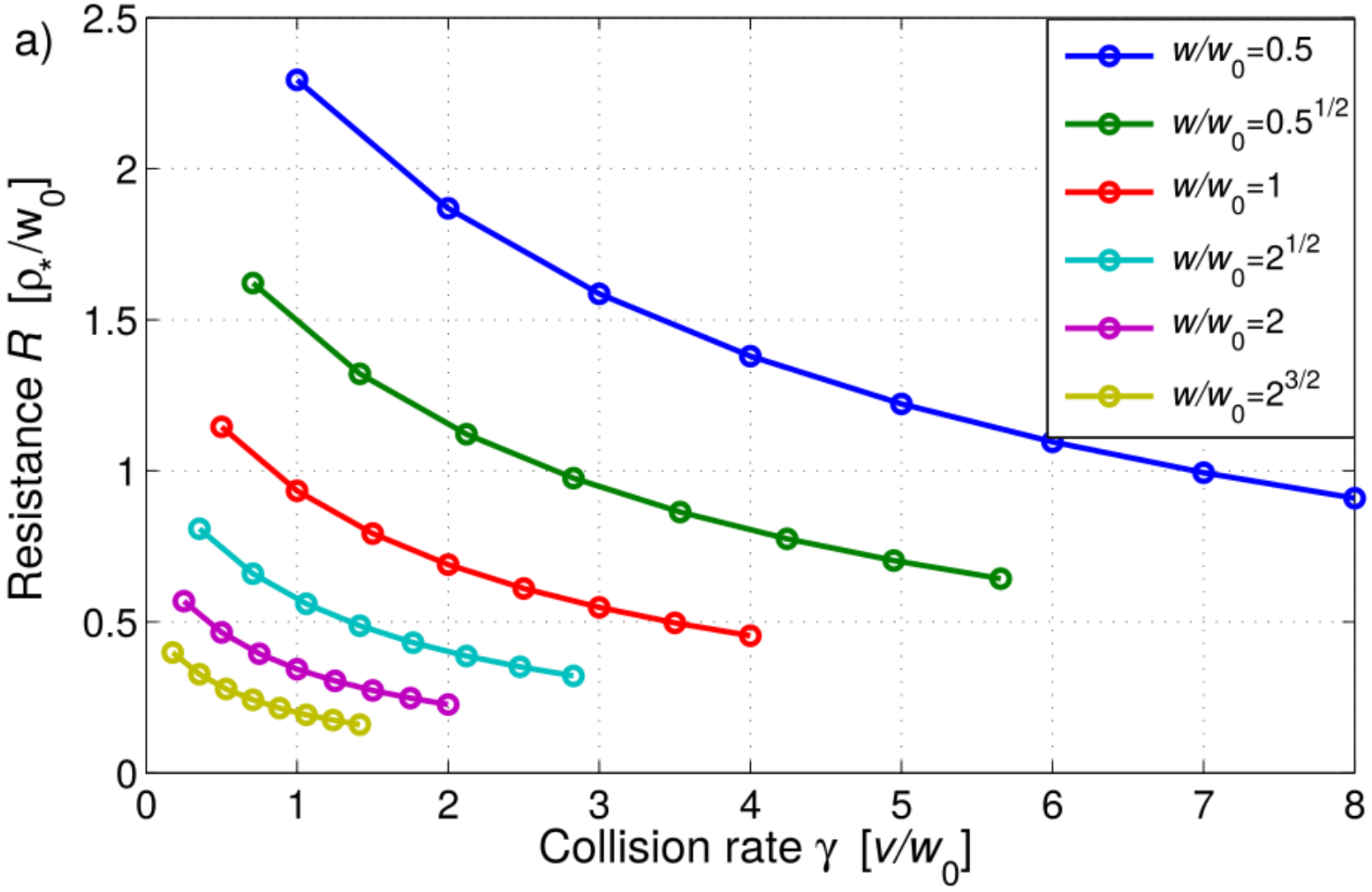} 
  \includegraphics[width=0.4\textwidth]{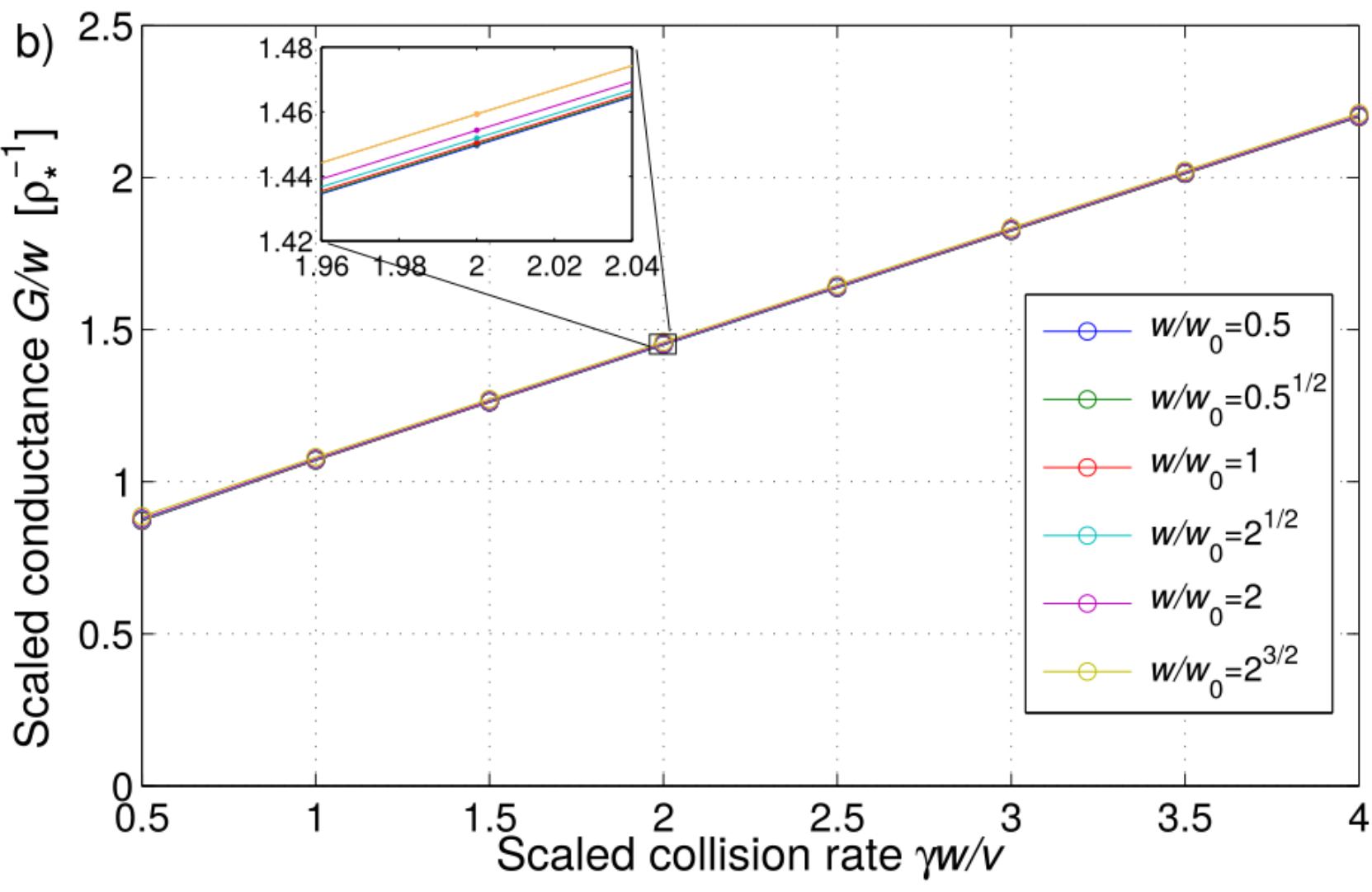} 
 \caption{a) 
The resistance $R$, Eq.\eqref{eq:R}, 
 plotted vs. $\gamma$. Upon rescaling 
 $R\to Rw$, $\gamma\to\gamma w$ all the curves collapse on one curve, 
 confirming that the only relevant parameter is the ratio $w/{l_{\rm ee}}=w\gamma/v$. 
 b) Scaled conductance 
 $G=1/(Rw)$ vs. $\gamma w$. 
All curves collapse onto a single straight line, which can be fitted with $(0.694 + 0.378 \gamma w)\rho_*^{-1}$. 
 Parameters used: 
 $b=10^6v$, the number of sampling points within the constriction $\sim160$, the length unit $w_0=\frac1{30}L$.
 }\label{fig3}
\end{figure}

We obtain current distribution by solving numerically Eq.\eqref{eq:line_y=0}, discretized as in Eq.\eqref{eq:discrete_k}, and subsequently Fourier-transforming $g_k$ to position space. 
A large value $b=10^5v$ was used to ensure that current vanishes outside the interval $[-{\frac{w}2},{\frac{w}2}]$. 
The resulting distribution, shown in Fig.\ref{fig1}b,
features interesting evolution under varying $\gamma$: 
flat at small $\gamma$, 
the distribution gradually bulges out as $\gamma$ increases, 
peaking at $x=0$ and dropping to zero near $x=\pm{\frac{w}2}$. In the 
limit $\gamma\gg v/w$ it evolves into a semicircle coinciding with the hydrodynamic result, Eq.\eqref{eq:semicircle}.
Current suppression near the constriction edges is in agreement with the streaming picture discussed above.

The solution on the line $y = 0$ can now be used to determine the solution in the bulk.
E.g. to obtain the density 
$f_0(\vec x)$ we project the relation \eqref{eq:(1+Ga)f=f0} on $m=0$ harmonic, taking into account that both $f^{(0)}$ and $\alpha \tilde f$ are of an $\ket{s}$ form. This allows to express the 2D density as 
$f_0(\vec x)=-\int dx' \tilde G_{0s}(\vec x,x')\alpha(x') g(x')$, with $\vec x$ a 2D coordinate and $-\infty<x'<\infty$.  
To avoid handling the $b\to\infty$ limit in $\alpha$, we write this relation using Eq.\eqref{eq:line_y=0} as
\be
f_0(\vec x)=-\int dx' \tilde G_{0s}(\vec x, x')\lp\mu-\lp D_{ss}^{-1} g\rp (x')\rp
.
\ee
Plugging $\tilde G_{0s}(\vec k)=\frac{-i\sqrt{2}k_2}{v(k_1^2+k_2^2)}$, Fourier-transforming, and carrying out the $k_2$ integral by the residue method, 
$
\int dk_2 
e^{ik_2 y}\frac{ik_2}{k_1^2+k_2^2}=-\pi 
e^{-|yk_2|}\sgn y 
$,
we obtain
\be\label{eq:f0_2D}
f_0(\vec x)=\frac{\sgn y}{\sqrt{2}v} \int \frac{dk_1}{2\pi} e^{ik_1x-|k_1y|}\lp  D_{ss}^{-1}(k_1) g_{k_1}-2\pi\mu\delta(k_1)\rp
.
\ee
The resulting distributions, shown in Fig.\ref{fig2}, are step-like. 
At large $y$ the $\mu$-term dominates, giving $f_0(|\vec x|\gg w)\approx -\frac{\mu}{\sqrt{2}v}\,\sgn y$. Therefore, the step height equals $\frac{\sqrt{2}}{v}\mu$ regardless of the parameter values used. 

This relation provides a route to evaluate 
resistance. Namely, because of 
charge neutrality, the density $f_0$ obtained from a noninteracting model translates directly into potential distribution $\phi(\vec x)=\frac1{e\nu_0} f_0(\vec x)$, where $\nu_0$ is the density of states. Dividing the potential difference $V=\frac{\sqrt{2}\mu}{e\nu_0 v}$ by
the total current $I=\int dx g(x)\braket{  ev\sin\theta}{s}
=\frac{ev}{\sqrt{2}}g_{k_1=0}$, yields a simple expression for resistance
\be\label{eq:R}
R=
\frac{\mu\rho_*}{vg_{k=0}}
,\quad
\rho_*=\frac2{e^2v\nu_0}
,
\ee 
where $g_{k=0}=\int  g(x)dx$ and $\rho_*$ is a constant of dimension ${\rm Ohm\cdot cm}$. 
Since $g\propto\mu$, the resulting $R$ values are $\mu$-independent. 
Fig.\ref{fig3}a shows $R$ plotted vs. 
$\gamma$. As expected, $R$ decreases as $\gamma$ increases, i.e. carrier collisions enhance conduction. 

As a quick sanity check on Eq.\eqref{eq:R} we consider the near-collisionless limit  $\gamma\ll v/w$. In this case $D_{ss}(k)\approx 2/\pi v$ and the integral equation \eqref{eq:line_y=0} turns into an algebraic equation which is solved by a step-like distribution
\be
g(|x|>w/2)=\frac{2\mu}{\pi v+ 2b}
,\quad
g(|x|<w/2)=\frac{2\mu}{\pi v}
.
\ee
In the limit $b\to \infty$ the total current is $I=\frac{ev}{\sqrt{2}}\frac{2w\mu}{\pi v}$. Taking the 2D density of states $\nu_0=\frac{Nm}{2\pi\hbar^2}$ (here $N$ is spin-valley degeneracy,  e.g. $N=4$ for graphene), we find 
\be
R=\frac{V}{I}=\frac1{N}\frac{h}{e^2}\frac{\lambda_F}{2w}
,\quad
\lambda_F=\frac{2\pi}{k_F}
\ee
This is precisely the collisionless Landauer value. 
Spatial dependence can be obtained 
by plugging $g(x)$ in Eq.\eqref{eq:f0_2D}. Integrating and taking the limit $b\to \infty$ gives
\be\label{eq:f0_ballistic}
f_0(\vec x)=-\frac{\sgn y}{\sqrt{2}v} \mu \lp 1-\frac1{\pi}\theta(\vec x)\rp
,
\ee
where $\theta(\vec x)= 
\tan^{-1}\frac{|y|w}{x^2+y^2-\frac14 w^2}$ is the angle at which  the interval $[-{\frac{w}2},{\frac{w}2}]$ is seen from the point $\vec x=(x,y)$. This confirms the result $\frac{\mu}{\sqrt{2}v}$ for the step height. 

The dependence $R$ vs. $\gamma$ 
shows several interesting features, some expected and some unexpected.
First, on general grounds,  we expect that the dependence on $\gamma$ is controlled solely by the ratio $w/{l_{\rm ee}}$. 
Indeed, plotting the rescaled quantity $R w$ vs. $\gamma w$ we find a family of curves that all collapse on one curve. 
Second, quite remarkably, inverting this quantity and plotting $1/(R w)$ vs. $\gamma w$ we find {\it a nearly perfect straight line} with a positive offset at $\gamma=0$, see Fig.\ref{fig3}b. The straight line, which is identical for all $w$ values, is described by 
$\rho_*/(R w)=a_1+a_2\gamma w$.  This dependence translates into a simple 
addition rule for 
conductance, $G=G_{\rm ball}+G_{\rm vis}$. 
The term $G_{\rm ball}$
describes a $\gamma$-independent ballistic contribution that scales linearly with $w$, whereas 
$G_{\rm vis}$ describes a viscous contribution proportional to $\gamma$ that scales as $w^2$. 
The two terms yield values $a_1=2/\pi$ and $a_2=\pi/8$, respectively. 
This is in good agreement with the values $a_1=0.694$, $a_2=0.378$ obtained from a best fit to the data in Fig.\ref{fig3}b.  

The additive behavior of conductance at the ballistic-to-viscous crossover comes as a surprise and, to the best of our knowledge, is not anticipated on simple grounds. 
This is in a stark departure from the Matthiessen's rule 
that 
mandates an additive behavior for resistivity in the presence of different scattering mechanisms, as observed in many solids\cite{matthiessen_rule}. This  rule is of course not valid if the factors affecting transport depend on each other, because individual scattering probabilities cannot be summed unless they are mutually independent. This is precisely the case for momentum-conserving ee collisions that do not by themselves result in momentum loss, but can impact momentum relaxation due to other scattering mechanisms. 
Furthermore, the 
addition rule for conductance, Eq.\eqref{eq:G=G+G}, describes a striking ``anti-Matthiessen'' behavior: rather than being suppressed by collisions, 
conductance exceeds the collisionless value.

\section{Supporting Information for ``Higher-Than-Ballistic Conduction of Viscous Electron Flows'' by Haoyu Guo, Ekin Ilseven, Gregory Falkovich and Leonid Levitov}

\section{A: Integral equation on a circle} 

The integral equation \eqref{eq:line_y=0}, 
which describes current distribution in the constriction, is defined on a line $-\infty<x<\infty$ in position representation. 
It reads
\bea\label{eq:integral_eqn_f(x)}
&&\alpha(x)g(x)+\int_{-\infty}^\infty dx' D_{ss}^{-1}(x-x')g(x') dx'=\mu
,\quad
\\ \nonumber
&& D_{ss}^{-1}(x-x')=\int\frac{dk}{2\pi} \frac{e^{ik(x-x')}}{D_{ss}(k)}
,\quad
\alpha(x)=\left\{\begin{array}{cc} b, & |x|>\frac{w}2 \\ 0, & |x|<\frac{w}2 \end{array}\right.
\eea
Before we proceed to discuss the general solution, 
it is instructive to consider Eq.\eqref{eq:integral_eqn_f(x)} in the collisionless limit ${l_{\rm ee}}\gg w$ and in the hydrodynamic limit $w\gg{l_{\rm ee}}$. These regimes are described by the large-$k$ and small-$k$ limits of $D_{ss}(k)$, given in Eq.\eqref{eq:Fss_asymptotic}. 

In the first case, $\gamma=0$ and $D_{ss}(k)= 2/\pi v$, and the integral equation \eqref{eq:integral_eqn_f(x)} turns into an algebraic equation. This equation is solved by
\be
f(|x|>w/2)=\frac{\mu}{1+\tilde b}
,\quad
f(|x|<w/2)=\frac2{\pi}\mu
.
\ee
where $\tilde b=2b/\pi v$. Taking the limit $b\to\infty$, describing a nontransparent boundary, we obtain a box-like solution that vanishes outside the slit $|x|<w/2$, which agrees with the current distribution in the ballistic limit $\gamma=0$.

In the second case, $\gamma\gg v/w$ and $D_{ss}(k)= \frac{\gamma}{|k|v^2}$, we have 
\bea\nonumber
&& D_{ss}^{-1}(x-x')=\int_{-\infty}^\infty\frac{dk}{2\pi} \frac{|k|v^2}{\gamma}e^{ik(x-x')}
\\ \label{eq:D=1/(x-x')^2}
&& =-\frac{v^2}{2\pi\gamma}\lp \frac1{(x-x'+i0)^2}+\frac1{(x-x'-i0)^2}\rp
\eea
This coincides with the kernel in Eq.\eqref{eq:int_eqn} of the main text. We will now show that the integral equation \eqref{eq:integral_eqn_f(x)}, in the limit $b\to\infty$, is satisfied by a semicircle solution identical to that found by an electrostatic method. 
The analysis is facilitated by representing the semicircle solution, with a yet-undetermined normalization factor, as
\[
g(x)=a\sqrt{1-\frac{4x^2}{w^2}}=\Im f_+(z)-\Im f_-(z)
,\quad
z=\frac{2x}{w},
\]
where $f_\pm(z)$ are given by $\frac{a}2(\sqrt{z^2-1}-z)$ continued from large $z$ to $-1<z<1$ through the upper or lower complex $z$ halfplane, respectively.  
Using this representation and the expression in Eq.\eqref{eq:D=1/(x-x')^2}, we can carry out the integral in Eq.\eqref{eq:integral_eqn_f(x)} by the method of residues, closing the integration path through the upper halfplane for $f_+(z)$ and the lower halfplane for $f_-(z)$. The contributions of large $z$ drop out since the functions $f_\pm(z)$ vanish at infinity, giving
\bea\nonumber
&& \int_{-\infty}^\infty dx' D_{ss}^{-1}(x-x')g(x') dx'
\\
&& =
\Im \lb \frac{2iv^2}{\gamma w}(f'_+(z)-f'_-(z))\rb
=\frac{2v^2a}{\gamma w}
.
\eea
Here we have taken $x$ to be in the interval $[-\frac{w}2,\frac{w}2]$. 
Inserting this result in Eq.\eqref{eq:integral_eqn_f(x)} we determine the normalization factor
$a=\frac{\gamma w}{2v^2}\mu$.
The resistance is obtained by evaluating 
\[
g_{k=0}=\int_{-\frac{w}2}^{\frac{w}2} g(x)dx=\frac{\pi}4 w a
\]
and plugging it in Eq.\eqref{eq:R}. This gives
\[
R=\frac{8v\rho_*}{\pi\gamma w^2}=\frac{16}{\pi e^2\gamma w^2\nu_0}
.
\]
Writing the 2D density of states as $\nu_0=\frac{Nm}{2\pi\hbar^2}=\frac{2n}{mv^2}$ and expressing $\gamma$ through viscosity $\eta=nmv^2/4\gamma$ [see Eq.\eqref{eq:viscous_mode}] we find
\[
R=\frac{8mv^2}{\pi e^2\gamma w^2n}=\frac{32\eta}{\pi e^2n^2w^2}
.
\]
This is precisely the hydrodynamic result given in Eq.\eqref{eq:R_vis}.

Next, to facilitate numerical analysis, we put our 2D problem on a cylinder, choosing a large enough cylinder circumference $L$ to provide a good approximation to the 2D problem. Closing the $x$ axis into a circle does not impact in any way the ${\rm 2D\to 1D}$ reduction, which yields an integral equation defined in the domain $[-\frac12L, \frac12L]$: 
\bea\label{eq:integral_eqn_f(x)_L}
&&\alpha(x)g(x)+\int_{-L/2}^{L/2} dx' \tilde D_{ss}^{-1}(x-x')g(x') dx'=\mu
,\quad
\\ \nonumber
&& \tilde D_{ss}^{-1}(x-x')=\sum_{m=-\infty}^\infty D_{ss}^{-1}(x-x'-mL)
\eea
with periodic boundary conditions, $g(x\pm L)=g(x)$. It may seem that the problem defined by Eq.\eqref{eq:integral_eqn_f(x)_L} is identical to that in Eq.\eqref{eq:integral_eqn_f(x)}, since any function $g(x)$ satisfying Eq.\eqref{eq:integral_eqn_f(x)_L}, after being continued periodically outside the domain $[-\frac12L, \frac12L]$, 
satisfies also Eq.\eqref{eq:integral_eqn_f(x)}. We note, however, that such a procedure generates functions which are nonzero not only in the constriction interval $[-{\frac{w}2},{\frac{w}2}]$ ($w<L$) but also in the intervals $[Lm-{\frac{w}2},Lm+{\frac{w}2}]$ 
where the solution of the original problem, Eq.\eqref{eq:integral_eqn_f(x)}, must vanish in the limit $b\to\infty$. Physically, this is equivalent to replacing one slit with an infinite array of slits of width $w$ each, and periodicity $L$. The behavior near one slit will not be affected by other slits so long as $L\gg w$. In our numerical study, taking $L$ equal few times $w$ was found sufficient to provide a reasonably good approximation. 

To handle the $L$-periodic boundary conditions, 
we write Eq.\eqref{eq:integral_eqn_f(x)_L} in momentum space, with momentum taking discrete values
\be\label{eq:k=2*pi*n/L}
k=\frac{2\pi n}{L}
\ee
where $n$ is an integer. 
We transform Eq.\eqref{eq:integral_eqn_f(x)_L} by 
inserting a resolution of identity $\frac1{L}\sum_k \ketbra{k}{k}=1$, and using $\braket{x}{k}=\exp(ikx)$, $D\ket{k_x}=D_{ss}(k_x)\ket{k_x}$, and $\braket{k}{g(x)}=g_k$, where
 \be
 g_k=\int_{-L/2}^{L/2} dx e^{-ikx} g(x)
 .
 \ee
 Finally, we obtain:
 \begin{equation}\label{eq:line2}
    \sum_{k_1}\alpha_{k-k_1}g_{k_1}+D_{ss}^{-1}(k)g_k=\mu L \delta_{k,0},
 \end{equation}
 where $\alpha_k=b(\delta_{k,0}-\frac{w}{L}\text{sinc}(kw/2))$.
 In numerical calculation, the values $n$ in Eq.\eqref{eq:k=2*pi*n/L} are limited by $-\frac12N\le n< \frac12 N$, where $N$ is a suitably chosen large number. This corresponds to discretizing functions $f(x)$ in position space by using an $N$-point mesh $x_i=\frac{iL}{N}$, $i=-\frac{N}2, -\frac{N}2+1, ..., \frac{N}2-1$ 
in the interval $[-\frac12L, \frac12L]$. 

We solve Eq.\eqref{eq:line2} numerically to obtain current distributions pictured in Fig.\ref{fig1}b. 
This was done by first finding the distribution $f_k$ in momentum space, and then Fourier-transforming to position space. We used $L=3w$, and a large value $b=10^5v$ to ensure that current vanishes outside the interval $|x|<{\frac{w}2}$. A Fourier space filter was used to smooth out the Gibbs phenomenon near the points $x=\pm {\frac{w}2}$ where current distribution drops abruptly to zero. 

In the plots the value $\mu$ was chosen such that the net current is normalized to unity. The resulting current distribution evolves in an interesting way upon $\gamma$ increasing: the distribution is a flat step at small $\gamma$, as expected in the ballistic case, and then gradually bulges forming a peak at $x=0$ and gradually dropping to zero near $x=\pm{\frac{w}2}$. In the extreme hydrodynamical limit $\gamma\gg v/w$, it evolves into a semicircle, which coincides with the result obtained from hydrodynamic equations in the main text.

\begin{figure}
  \includegraphics[width=0.4\textwidth]{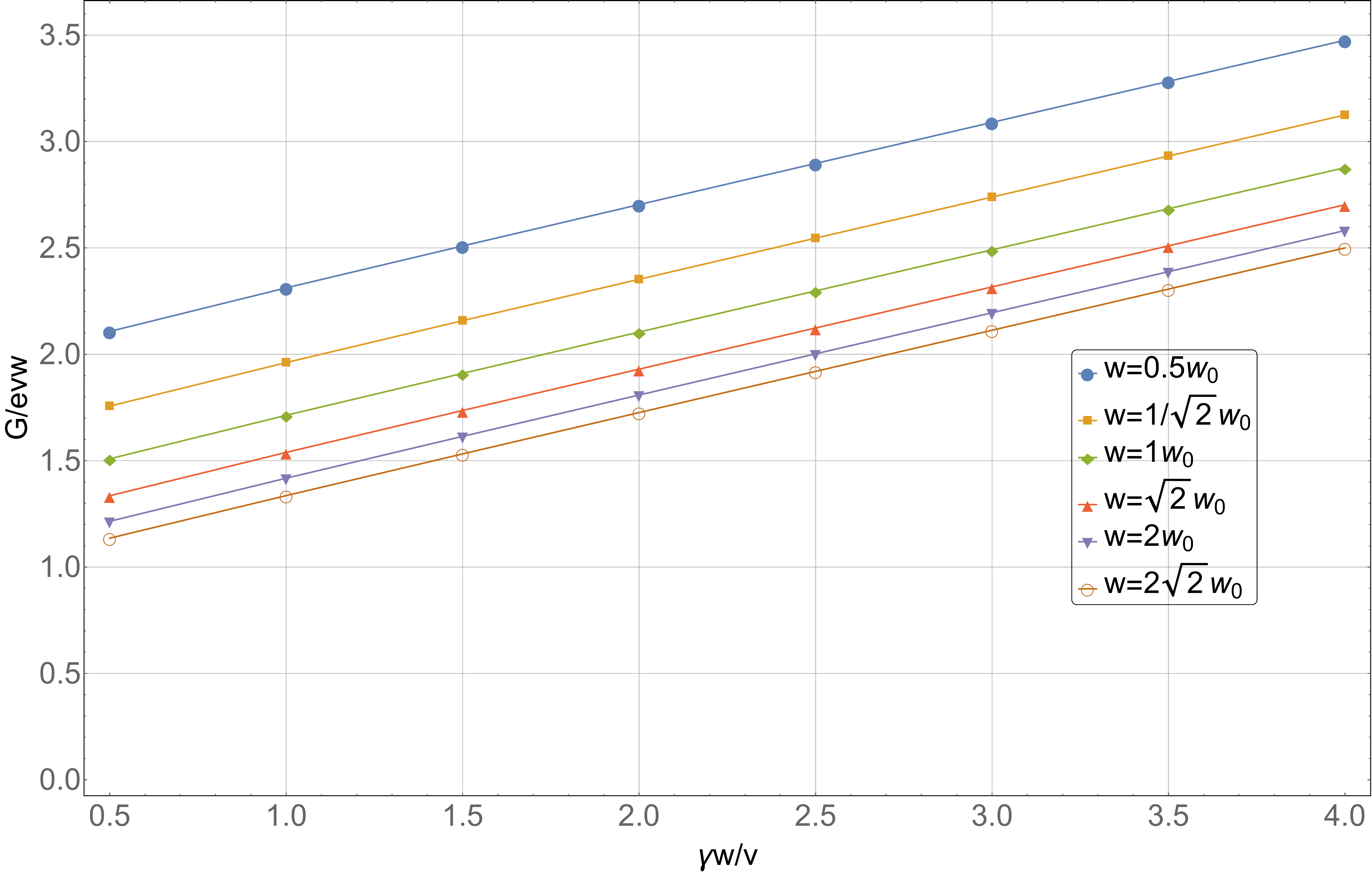}
  \caption{Conductance per width vs. $\gamma w$. Plots are obtained at $w_0=\frac1{30}L$, $b=50v$, with the number of sampling points within the constriction of about $160$. Unlike Fig.\ref{fig3}, here different curves do not collapse on one curve, indicating that the universality fails for not-too-large $b$.  }\label{fig6}
\end{figure}

Using the solution $g_k$, resistance $R$ can be calculated from Eq.\eqref{eq:R}, giving the conductance $G=1/R$ shown in Fig.\ref{fig3} and Fig.\ref{fig6}. For large $b=10^6v$, the conductance plots $G$ vs. $\gamma$, obtained for different constriction widths $w$, collapse on one curve when rescaled to $G/w$ vs. $\gamma w$. This `universality' confirms that the only relevant parameter in the problem is the ratio $w/{l_{\rm ee}}$. This scaling stops working already for not very large $b$, as illustrated in Fig.\ref{fig6}. The breakdown of scaling is not alarming, since physically meaningful results are expected only in the limit $b\to \infty$. Interestingly, however, the dependence  $G/w$ vs. $\gamma w$ is well fitted by a perfectly straight line both for $b$ large and not-too-large. The linear dependence $G$ vs. $\gamma$, along with the scaling, indicate that the conductance at the crossover is described by the addition formula $G=G_{\rm ball}+G_{\rm vis}$, as discussed in the main text.

\section{B: Extreme Hydrodynamic Regime}
    Here we use the approach developed in the main text to solve for the 2D potential distribution, current flow 
and conductance 
in the hydrodynamic regime $\gamma\gg v/w$. 
In this case, the solution of the integral equation \eqref{eq:integral_eqn_f(x)} is a semicircle 
\begin{equation}
    g(x)=a\sqrt{1-\frac{4x^2}{w^2}},
\end{equation}
    where $a$ is a normalization factor which for the time being we will leave undetermined. 
    In the  Fourier domain,
\begin{equation}
    g(k)=
    \int_{-\frac{w}2}^{\frac{w}2} g(x) 
    e^{-ikx} \rd{x}=
    \frac{\pi w a}2\frac{J_1(|kw/2|)}{|kw/2|},
\end{equation}
    where $J_1$ is the Bessel Function.
The solution of the 1D problem can be used to obtain the 2D flow by using the same procedure as the one employed to obtain the density distribution, Eq.\eqref{eq:f0_2D}. 
Using the values $\tilde G_{cs}$ and $\tilde G_{ss}$ given in Eq.\eqref{eq:Gtilde}, and approximating $D_{ss}^{-1}\approx |k_x| v^2/\gamma$, the $\ket{c}$ and $\ket{s}$ components of the flow are given by
%
\begin{equation}
    f(k_x,k_y)=
    \frac{4\pi a}{v^2} J_1(|k_xw/2|)\left(
                                                \begin{array}{c}
                                                  \frac{-k_xk_y}{k^4} \\
                                                  \frac{k_x^2}{k^4} \\
                                                \end{array}
                                              \right),
\end{equation}
where the two entries represent the $x$ and $y$ momentum components, respectively. In the equation above, the $\mu$ term does not contribute.
    The next step is to perform Fourier transform to obtain the real-space flow distribution $f(x,y)=\int\frac{\rd{k_x}}{2\pi}\frac{\rd{k_y}}{2\pi}f(k_x,k_y)e^{ik_xx+ik_yy}$.
 After the $k_y$ integral is calculated by the residue method, we have
\be
\begin{split}
   & f(x,y)=\int\frac{\rd{k_x}}{2\pi}\frac{
   \pi a e^{ik_xx-|k_xy|}}{v^2}J_1(|k_xw/2|)\left(
                                                                              \begin{array}{c}
                                                                                -iy \\
                                                                                \frac{1+|k_xy|}{|k_x|} \\
                                                                              \end{array}
                                                                            \right) \\
   & =\Re \int_{0}^{\infty}\rd{k_x} \frac{
   a e^{ik_xx-|k_xy|}}{v^2}J_1(|k_xw/2|)\left(
                                                                              \begin{array}{c}
                                                                                -iy \\
                                                                                \frac{1+|k_xy|}{|k_x|} \\
                                                                              \end{array}
                                                                            \right)
\end{split}
\ee
The $k_x$ integral can be evaluated using the identity 
\be
    \int_0^{\infty}e^{-\alpha x}J_\nu(\beta x)=\frac{\beta^{-\nu}(\sqrt{\alpha^2+\beta^2}-\alpha)^\nu}{\sqrt{\alpha^2+\beta^2}}. 
\ee
This gives the flow velocity components:
\be
\begin{split}
    f_x&=
    -\frac{2 a}{w v^2} \Im \lp y\frac{|y|-ix}{Z}\rp,\\
    f_y&=
    \frac{2a}{w v^2} \Re \lp Z-|y|\frac{|y|-ix}{Z}\rp,\\
    Z&=\sqrt{(w/2)^2+(|y|-ix)^2}.
\end{split}
\ee
    The resulting flow is shown in Fig.\ref{fig1}. Using Eq.\eqref{eq:f0_2D} we can compute the density distribution: 
\begin{equation}
 f_0(x,y)=\frac{\sgn y}{\sqrt{2}v}\lp 
 \frac{2 a}{\gamma w} \Re\lp 1-\frac{|y|-ix}{Z}\rp -\mu\rp 
 .
\end{equation}
    The value of $\mu$ is determined by $f_0$ continuity at the constriction, giving $\frac{1}{\mu w}=(\pi/8) \gamma w\sim 0.39\gamma w$. This agrees with the analysis in the main text: 
    the numerical results plotted in Fig.\ref{fig3} yield the best-fit slope $\sim 0.378$, which is in good agreement with our analytic results.

\section{C:
Hydrodynamic modes}

Here we derive hydrodynamic modes using the method of quasi-hydrodynamic variables, developed in the main text. This will allow us to relate the collision rate $\gamma$ and viscosity. To that end we consider Eq.\eqref{eq:Boltzmann} in the absence of boundary scattering, $I_{\rm bd}=0$.
In this case, Eq.\eqref{eq:Boltzmann} takes the form
\be\label{eq:(L-gamma*P)f=0}
(\hat K-\gamma P)f=0
,\quad
\hat K=\p_t+\vec v\nabla_{\vec x}+\gamma \hat 1
.
\ee
Since $f_0$ and $f_{\pm1}$ are zero modes of the collision operator  $I_{\rm ee}$, they dominate at low frequencies and long wavelengths. Accordingly, we can obtain hydrodynamic modes
from plane-wave solutions, $f(\theta,\vec x,t)\sim f(\theta) e^{-i\omega t+i\vec k\vec x}$. Solving  Eq.\eqref{eq:(L-gamma*P)f=0} as $f=\gamma \hat K^{-1}Pf$ we project $f$ on  the harmonics $f_0$ and $f_{\pm1}$. This gives three coupled equations
\be
f_m=g_{mm'}f_{m'}
,\quad
g_{mm'}=\bra{ m}\gamma P\hat K^{-1}P\ket{m'}
.
\ee
Direct calculation gives
\be\label{eq:g3x3_appendix}
g_{mm'}=\la \frac{\gamma e^{i(m'-m)\theta}}{\gamma_\omega+i\vec k\vec v}\ra_\theta
\!\!
=\tanh\beta \frac{\gamma e^{i\theta_k\Delta m} }{\gamma_\omega\lp ie^{\beta}\rp^{|\Delta m|}}
,
\ee
where $\la ...\ra_\theta=\oint ...\frac{\theta}{2\pi}$. 
Here $\gamma_\omega=\gamma-i\omega$, $\sinh\beta=\frac{\gamma_\omega}{kv}$  and $\Delta m=m'-m$, $m,m'=0,\pm 1$. The integral over $\theta$ in Eq.\eqref{eq:g3x3_appendix} is performed by writing
$\vec k\vec v=kv\cos\tilde\theta$, where $\tilde\theta=\theta-\theta_k$ is the angle between particle velocity $\vec v$ and wavevector $\vec k$, and integrating over $\tilde\theta$. 

As we now show, the equations
$f_m=g_{mm'}f_{m'}$
generate an acoustic and a viscous mode. Since the acoustic and viscous modes are longitudinal and transverse, respectively, it is convenient to do the analysis by 
performing an orthogonal transformation to the even/odd basis
\be
\left.|0\ra,\quad \left.|c\ra=\frac{\left.|1_k\ra+\left.|-1_k\ra}{\sqrt{2}}
,\quad
\left.|s\ra=\frac{\left.|1_k\ra-\left.|-1_k\ra}{\sqrt{2}i}
,
\ee
where we use notation $\left.|m_k\ra=e^{-im\theta_k}\left.|m\ra$.
The modes $\left.|c\ra$ and $\left.|s\ra$
correspond to normalized angular harmonics $f_c(\theta)=\sqrt{2}\cos \tilde\theta$ and $f_s(\theta)=\sqrt{2}\sin \tilde\theta$.

This transformation brings  the $3\times 3$ matrix $g_{mm'}$ to a block-diagonal form
\be
\lp\begin{array}{ccc} g_{00} & g_{0c} & 0 \\ g_{c0} & g_{cc} & 0 \\ 0 & 0 & g_{ss} \end{array}\rp
.
\ee 
For the odd-mode $1\times 1$ block we find $g_{ss}=\frac{\gamma}{\gamma_\omega}\tanh\beta (1+e^{-2\beta})$. Writing the dispersion relation $1=g_{ss}$ and Taylor-expanding in small $\omega$ and $k$ yields a viscous mode dispersing as
\be\label{eq:viscous_mode}
\omega=-i\nu k^2
,\quad
\nu=v^2/4\gamma
.
\ee
Here $\nu$ is the viscosity defined so that the dispersion in Eq.\eqref{eq:viscous_mode} agrees with that obtained from the linearized Navier-Stokes equation $(\p_t-\nu\nabla^2)\vec v=-\nabla P$.

The acoustic mode can be obtained from the even-mode $2\times2$ block 
\be
\lp\begin{array}{cc} g_{00} & g_{0c} \\ g_{c0} & g_{cc}\end{array}\rp
=
\frac{\gamma\tanh\beta}{\gamma_\omega}
\lp\begin{array}{cc} 1 & -i\sqrt{2} e^{-\beta} \\  -i\sqrt{2} e^{-\beta} & 1-e^{-2\beta}\end{array}\rp
.
\ee
The dispersion relation ${\rm det}\,(1-g)=0$ gives
\be
\lp \frac{\gamma_\omega}{\gamma\tanh\beta}-1\rp \lp \frac{\gamma_\omega}{\gamma\tanh\beta}-1+e^{-2\beta}\rp +2e^{-2\beta}=0
\ee
Plugging $\sinh\beta=\frac{\gamma}{kv}$, simplifying and Taylor-expanding in  $\omega$ and $k$, yields a damped acoustic mode
\be
\omega=\frac1{\sqrt{2}}kv-\frac{i}2\nu k^2
,
\ee 
where we expressed damping through viscosity $\nu$, evaluated in Eq.\eqref{eq:viscous_mode}.

\end{document}